\documentclass[aip,pop,preprint,superscriptaddress,floatfix]{revtex4-2}
\usepackage[caption=false]{subfig} 
\usepackage[hidelinks]{hyperref}   
\hypersetup{
  colorlinks = true,               
  urlcolor   = blue,               
  linkcolor  = blue,               
  citecolor  = blue,               
  breaklinks = true                
}
\usepackage{siunitx}               
\usepackage{bm}                    
\usepackage[version=4]{mhchem}     
\usepackage{booktabs}              

\newcommand{\ifrac}[2]{{#1}/{#2}}
\newcommand*\diff{\mathop{}\!\mathrm{d}}
\newcommand{\fpartial}[2]{\frac{\partial {#1}}{\partial {#2}}}

\newcommand{\ipartial}[2]{\ifrac{\partial {#1}}{\partial {#2}}}

\newcommand{\ie}{\emph{i.e.}}

\newcommand{\rom}[1]{\uppercase\expandafter{\romannumeral #1\relax}}
\renewcommand{\vec}{\mathbf}
\DeclareMathOperator\erf{erf}

\newcommand{\UPM}{Department of Applied Physics. Escuela T\'{e}cnica Superior de Ingenier\'{i}a Aeron\'{a}utica y del Espacio. Universidad Polit\'{e}cnica de Madrid, 28040 Madrid. Spain}

\graphicspath{{figures/}}

\begin{document}
  \title{On the particle and electromagnetic boundary conditions in a mesothermal plasma flow}
  \author{J.~Gonzalez}
  \affiliation{\UPM}

  \date{\today}
  \begin{abstract}
    The fully kinetic modeling of mesothermal plasma flows presents significant challenges in the description of boundary conditions for the electrons and the electromagnetic field.
    Due to their higher mobility, electrons extend downstream of the ions, which hinders the implementation of the outflow boundary conditions during transient and quasi-steady states.
    The electron drift velocity, $u_e$, is usually neglected as $u_e \simeq u_i < v_{th,e}$, where $u_i$ is the ion bulk velocity and $v_{th,e}$ is the electron thermal velocity.
    However, we will show it is important to account for it at the injection of particles and at the outflow boundary to achieve quasi-neutrality of charge and current densities.
    A new partial reflection model for electrons is presented to represent quasi-steady and transitory states.
    Additionally, we will show how we can employ Ampere-Maxwell's law to better represent the transitory electric field at the outflow boundary compared to the classical `zero-field' Neumann boundary and without relying in non-physical numerical parameters.
    With these new boundary conditions, we improve the quasi-steady and transitory descriptions of expanding mesothermal plasmas.
  
  \end{abstract}
  
  \maketitle
  \section{Introduction}
    The expansion of plasma into a vacuum covers situations as diverse as the expansion of a high-Z high-density plasma~\cite{murakami2005ion, bochkarev2006investigation, doggett2011expansion} or the wake of low-density interplanetary plasma.~\cite{cai2015numerical, wang2018breakdown, shi2022acceleration}
    Of particular interest is the expansion of mesothermal plasmas, \ie, a flowing plasma in which the module of the ion drift velocity, $u_i$, is larger than their thermal velocity, $v_{th,i} = \sqrt{\ifrac{2 q T_i}{m_i}}$, but way smaller than the electron thermal velocity, $v_{th,e} = \sqrt{\ifrac{2 q T_e}{m_e}}$, typically expressed as $v_{th,i} < u_i \ll v_{th,e}$, where $T_i < T_e$, and $m_i \gg m_e$ are the temperature and mass of the ions and electrons, respectively, and $q$ is the elementary charge.
    Ions move in a preferred direction, typically coincident with the thruster axis, at a drift velocity of $u_i$ with very low dispersion, while the perpendicular dynamic is determined by $v_{th,i}$ and the self-consistent electric field.
    On the other hand, electrons move much faster due to their lower mass and higher temperature, which makes them progress further than the ions.
    This type of flow appears in space propulsion devices based on plasma,~\cite{roy1996ion, wang2015kinetic, hu2018expansion} in which a hypersonic plume is exhausted from the vessel to generate thrust.
    In these devices, the correct description of the plasma plume is important to understand processes such as the plasma backflow.~\cite{roy1996ion, cescon2026plume}
    
    During the expansion of the plume, a potential drop is formed at the ion front, \ie, the furthest point reached by the ions.
    As electrons are highly movable, they reach positions downstream of the ion front.
    A fraction of electrons is reflected by the potential drop, so the plasma upstream of the front remains quasi-neutral.
    This quasi-neutrality not only implies a zero charge density, $\sum \rho_j \simeq 0$, but also zero current density in the plasma, $\sum \vec{J}_j \simeq 0$, where $\rho_j = q_j n_j $ is the charge density, $\vec{J}_j = q_j \vec{u}_j n_j$ is the current density, $q_j$ is the species charge, $n_j$, is the species particle density, $\vec{u}_j$ is the drift velocity, and the subscript $j$ determines the species, typically electrons and ions.
    Of course, this requires that the electron drift velocity, $\vec{u}_e$ to be non-zero so that the current of positive and negative charges equalizes.
    
    Nevertheless, many models neglect the drift velocity of electrons at the injection~\cite{li2019electron, wang2012electron–ion, wang2015kinetic,hu2017fully} under the assumption that $u_e \ll v_{th,e}$ and thus the most typical velocity of an electron will be larger than the drift velocity.
    This needs a correction at the injection boundary,~\cite{li2019electron} as the injected current with a half-Maxwellian distribution, which assumes no drift, is not enough to maintain the quasi-neutrality of the expanding plasma.
    Additionally, when the ion front reaches the outflow boundary, the partial reflection of the electrons due to the potential drop must be accounted for, as well as maintaining quasi-neutrality of charge and current density in the quasi-steady state.
    
    Moreover, the boundary condition for the electric field $\vec{E}$ at the outflow boundary is not immediate.
    In plasma expansion models, this field is typically derived from the Poisson's equation, assuming an electrostatic approach
    \begin{equation}
      \nabla^2 \phi = - \frac{1}{\varepsilon_0} \sum \rho_j \label{eq:Poisson}
    \end{equation} 
     where $\nabla^2$ is the Laplace operator, $\phi$ is the electric potential such as $\vec{E} = -\nabla \phi$, and $\varepsilon_0$ is the vacuum permittivity.
    This is done even when an external fixed magnetic field is applied, as in the case of a magnetic nozzle.~\cite{andrews2022fully}
    The two typical boundary conditions for Eq.~\eqref{eq:Poisson} are a constant potential (Dirichlet), usually for metallic walls connected to power supplies, or setting the electric field (Neumann), which is typically zero when the boundary is considered `open'.~\cite{wang2015kinetic, hu2017fully, li2019electron}
    A combination of both conditions can be imposed in the so-called mixed (Robin) boundary condition.~\cite{brieda2018model}
    
    At the ion front, a strong charge density appears due to $\sum \rho_j \neq 0$, creating a non-zero electric field that must be accounted for when the front crosses the outflow boundary.
    Then, in the quasi-steady state, the electric field must be close to zero, as the plasma is quasi-neutral.
    If the dynamic of the electrons and the electric field close to the outflow boundary condition are not properly modeled, the so-called `numerical pump instability' appears.~\cite{brieda2018model}
    
    Authors in the past have proposed different ways to model the outflow boundary conditions when simulating a flowing plasma.
    For example, if the simulation is stopped before the plasma reaches the boundary, no pump instability appears.~\cite{hu2017fully, wang2018breakdown, wang2019limitations}
    However, this constrains the time that can be simulated, as well as not allowing the study of transitory effects.
    Increasing the simulation time requires of longer numerical domains, which increases the computational cost.
    
    Many authors employ employ a partial reflection model for the electrons at the outflow boundary, based on the potential difference between the boundary and injection plane,~\cite{li2019electron, andrews2022fully, difede2023magnetic, yang2024numerical, guaita2025electron, cescon2026plume} or the total charge in the numerical domain.~\cite{jambunathan2020self}
    These approaches capture the dynamics of electrons but present the issue of requiring information external to the boundary, which might hinder their implementation in complex domains.
    Moreover, the electric field is imposed as zero at the outlet boundary in many of these works,~\cite{li2019electron, yang2024numerical} which results in an incomplete description of the transient plasma.
    
    To avoid the pump instability, Brieda,~\cite{brieda2018model} used an `ion sink', a fictitious surface inside the numerical domain that absorbs ions but is transparent to electrons.
    This removes the need of imposing a boundary condition for the electrons and electric potential to represent the outflow of the plasma.
    However, this leaves a fraction of the numerical domain unusable, similar to what happens when zero electric field is imposed,~\cite{li2019electron} as the ion sink has to be far enough from the domain boundary to not affect the plasma dynamics.
    
    Recently, Andrews \textit{et al.}~\cite{andrews2022fully} proposed to treat the outflow boundary surface as electrically connected to ground, set at an infinite distance, by a capacitor, which has been implemented by different authors.~\cite{difede2023magnetic, guaita2025electron, cescon2026plume}
    The ground potential is updated at each iteration as
    \begin{equation}
        \phi_\infty^{k+1} = \phi_\infty^{k} + \frac{\tau}{C} \sum I_j^{k},
    \end{equation}
     where $I_j^{k} = q_j n_j^{k} \vec{u}_j^{k} \cdot \vec{S}$ is the current of charged species $j$ exiting a surface $\vec{S}$ at the iteration $k$, $\tau$ is the simulation time step, and $C$ is the `virtual free-space capacity'.
    Then, a non-stationary Robin boundary condition is used to relate $\phi_\infty$ to the potential at the boundary.
    The value of this free-space capacity affects the dynamics of the plasma close to the boundary and requires fine-tuning of the value for each specific case, which hinders its implementation as a general boundary condition.
    
    In this work we review the use of half-Maxwellian and Maxwellian distribution functions for the injection of electrons and how this affects the outflow boundary condition for the particles to maintain the quasi-neutrality of density and current in the plasma.
    Additionally, we present a new kind of boundary condition for the electric potential at the outflow boundary condition, based on Ampere-Maxwell's law, neglecting the effect of the magnetic field, that represents the dynamic of the electric field when the ion front crosses the boundary without the need for any numerical parameter.
    
    The remaining sections are organized as follows.
    Section~\ref{sec:numericalSetup} introduces the numerical setup employed for the simulations to test different boundary conditions.
    The injection of electrons with a half-Maxwellian distribution function and its consequences are studied in Sec.~\ref{sec:haflMaxwellian}.
    Then, in Sec.~\ref{sec:driftMaxwellian}, we inject the electrons with a drift velocity to match that of the ions and adjust the outflow reflection boundary condition so that quasi-neutrality of current is maintained in the quasi-steady state.
    In Sec.\ref{sec:Ampere}, we present a new boundary condition for the Poisson's equation based on the Ampere-Maxwell's law that represents the transition region in which the ion front crosses the boundary better than the classical zero-field Neumann one.
    Finally, conclusions are given in Sec.~\ref{sec:conclusions}.

  \section{The numerical setup}\label{sec:numericalSetup}
    To test the different combinations of particle and electric potential boundary conditions, we establish a common simulation case.
    The Finite element PArticle Kinetic Code (\textit{fpakc})~\cite{gonzalez2026fpakc} has been used for these simulations.
    This code is currently being developed to simulate mainly low-density plasmas from a fully-kinetic approach.~\cite{gonzalez2019particle, gonzalez2024self}
    The code carries out the typical steps in a particle-in-cell code: macro-particle pushing, with a Boris algorithm,~\cite{boris1970relativistic, birdsall2004plasma} scattering, field calculation, and gathering.
    The macro-particles, each one representing a significant number of real particles, are traced in an unstructured mesh, composed of elements that can be nodes (0D), segments (1D), triangles and rectangles (2D), and tetrahedrons (3D).
    
    \begin{figure}
      \centering
      \includegraphics{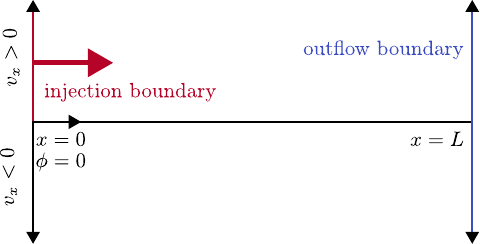}
      \caption{\label{fig:domainScheme}
               Scheme of the 1D-numerical domain employed in the simulation from $x = 0$ up to $x = L$.
               The remaining velocity components are not shown, but they are included in the particle push.
               Red line indicates the surface for injection of particles (inflow).
               Blue line represents the outflow boundary.
               A Dirichlet boundary condition for the Poisson's equation is imposed at $x=0$.
              }
    \end{figure}
    
    We employ a one-dimensional Cartesian domain, but due to the finite element nature of the mesh in the code, the implementation of different boundary conditions can be easily extended to higher dimensions, as all elements of the mesh are treated as abstract objects.
    As shown in Fig.~\ref{fig:domainScheme}, the domain extends from $0$ up to $L$ in the $x$ coordinate with its associated velocity $v_x$.
    The other spatial dimensions are considered infinite, and the velocity is always tracked in a 3D domain in fpakc.
    The plasma is injected at the $x = 0, v_x > 0$ region, and boundary conditions for the particles, either reflection or absorption, are imposed at $x = L = \qty{0.1}{\m}$.
    Ions are always removed from the domain when they reach the outflow boundary.
    The electric potential at $x = 0$ is set to $0$ to represent the plasma potential.
    At the outflow boundary, we will employ a `zero-field' Neumann, boundary condition, \ie, $\ipartial{\phi}{x} = 0$ for our 1D case, or the new boundary condition introduced in Sec.~\ref{sec:Ampere}.
    The mesh is composed of segments of $\Delta x = \qty{0.4}{\mm}$ in length, half the Debye length $\lambda_{D}$ at the injection conditions described below.
    The time step to advance both species in the simulation is $\tau = \qty{1e-11}{\s}$, which fulfills the classical limit in PIC simulations $\tau \omega_p < 0.1$,~\cite{brieda2005development,tskhakaya2007particle‐in‐cell} where $\omega_p = \sqrt{\ifrac{n_e q^2}{\varepsilon_0 m_e}}$ is the electron plasma frequency.
    
    We inject an argon plasma composed of single charged ions and neutrals, but for many of the models presented in sections~\ref{sec:driftMaxwellian} and~\ref{sec:Ampere} we will generalize to an indeterminate number of charged species.
    Ions are injected with a drift velocity of $u_{i,0} = \qty{40}{\km\per\s}$ and a temperature of $T_{i,0} = \qty{0.172}{\eV}$ (\qty{2000}{\K}) forming a narrow Maxwellian distribution.
    The density current for the injected ions is $J_{i,0} = \qty{1}{\ampere\per\square\meter}$, which gives a charge density at $x = 0$ of $\rho_{i,0} = \ifrac{J_{i,0}}{u_{i,0}} = \qty{25}{\uC\per\cubic\meter}$.
    Electrons have a temperature of $T_{e,0} = \qty{2}{\eV}$ and they are injected as half-Maxwellian distributions, Sec.~\ref{sec:haflMaxwellian}, or with a drift velocity equal to $u_{i,0}$, sections~\ref{sec:driftMaxwellian} and~\ref{sec:Ampere}.
    The density current injected for electrons necessary to maintain quasi-neutrality is defined as,
    \begin{equation}
        J_{e,0} = \rho_{i,0} \int\limits_{0}^\infty \frac{v_x}{\sqrt{\pi} v_{th,e}} \exp\left[ - \frac{\left(v_x - u_{e,0}\right)^2}{v_{th,e}^2}\right] \diff {v_x},
    \end{equation}
     where $v_{th,e} = \sqrt{\ifrac{2 q T_{e,0}}{m_e}}$ is the electron thermal velocity at the injection plane.
    After integration, the current becomes
    \begin{equation}
        J_{e,0} = \frac{\rho_{i,0}}{2} \left[\frac{v_{th,e}\exp\left(-\frac{u_{e,0}^2}{v_{th,e}^2}\right)}{\sqrt{\pi}} + u_{e,0} \erf\left(\frac{u_{e,0}}{v_{th,e}}\right) + u_{e,0} \right],
    \end{equation}
     which is larger than the ion injection current $J_{i,0} = \rho_{i,0} u_{i,0} $ as $v_{th,e} \gg u_{i,0} \gtrsim u_{e,0}$.
    For example, for a half-Maxwellian distribution function, \ie, $u_{e,0} = 0$, the ratio of injected currents becomes
    \begin{equation}
        \frac{J_{e,0}}{J_{i,0}} = \frac{v_{th,e}}{2 u_{i,0} \sqrt{\pi}} \simeq 5.92,
    \end{equation}
     for the values presented above.
    At this point, it is worth remembering that the injecting electron current only accounts for the positive velocity region, $v_x > 0$.
    Once electrons are reflected, a negative current will return to the injection plane so that the total electron current equals that of the injected ions.
    
    At each iteration of our simulations, $20$ electron and $5$ ion macro-particles are injected into the numerical domain.
    These particles have the corresponding statistical weight so that the equivalent current injected is the imposed one for each species.
    During the simulation, the weight of the injected electrons is adjusted to maintain the electric field close to zero.
    The value of the macro-particle weight should still oscillate around the one at the start of the simulation, as this correction is intended to compensate for the noise intrinsic to PIC simulations.
    At each iteration, the new weight of the electron macro-particles is updated as
    \begin{equation}
        W^{k+1} = W^{k} - K \vec{E} \cdot \vec{n}_b 
    \end{equation}
     where $W$ is the macro-particle weight, $k$ is the current iteration, $\vec{n}_b$ is the vector normal to the injection surface that points inwards to the numerical domain, and $K$ is a numerical factor to smooth the noise and avoid abrupt changes, typically set to \num{1e-1}.
    This correction ensures that the electric field at the injection boundary remains close to $0$.
     
    We define the time at which the ion front reaches the outflow boundary condition as $t_{B} = \ifrac{L}{u_{i,0}} = \qty{2.5}{\us}$.
    Simulation will continue up to $t_f = \qty{4}{\us}$ to analyze the quasi-steady behavior after the ion front has exited the simulation domain
    
    We have chosen a set of non-dimensional values for the simulations presented below, such as
    \begin{equation}
      \begin{gathered}
        \tilde{x}      = \frac{x}{L},     \;
        \tilde{t}      = \frac{t}{t_B},   \\
        \tilde{\rho}_j = \frac{\rho_j}{\rho_{i,0}},\;
        \tilde{u}_j    = \frac{u_j}{u_{i,0}}, \;
        \tilde{J}_j    = \frac{J_j}{J_{i,0}}, \\
        \tilde{E}      = \frac{E L}{T_{e,0}},
      \end{gathered}
    \end{equation}
     where the tilde indicates a non-dimensional value and the subscript $j$ represents a charged species, typically an ion or electron.
    
  \section{Injection of half-Maxwellian electrons and its consequences}\label{sec:haflMaxwellian}
    Many simulations inject electrons with a half-Maxwellian distribution, \ie, with no drift velocity.~\cite{li2019electron,wang2012electron–ion,wang2015kinetic,hu2017fully}
    This is mostly justified by assuming that $u_e \ll v_{th,e}$ and by the fact that the injection of particles following a Maxwellian distribution is much easier to implement than for a drifted Maxwellian.~\cite{cartwright2000loading}
    
    In the case presented here, we assume electrons have no drift-velocity, and thus they are injected with a half-Maxwellian distribution.
    At the outflow boundary, to keep the same condition as the injection, we impose a full reflection of electrons.
    Zero-field Neumann boundary condition for the Poisson's equation at $\tilde{x}=1$.
    
    \begin{figure*}
      \centering
      \renewcommand*{\thesubfigure}{\alph{subfigure}.1}
      \subfloat{\label{fig:HM_timeEvolution_rho_1}}
      \addtocounter{subfigure}{-1}
      \renewcommand*{\thesubfigure}{\alph{subfigure}.2}
      \subfloat{\label{fig:HM_timeEvolution_rho_2}}
      \addtocounter{subfigure}{-1}
      \renewcommand*{\thesubfigure}{\alph{subfigure}.3}
      \subfloat{\label{fig:HM_timeEvolution_rho_3}}
      \renewcommand*{\thesubfigure}{\alph{subfigure}.1}
      \subfloat{\label{fig:HM_timeEvolution_J_1}}
      \addtocounter{subfigure}{-1}
      \renewcommand*{\thesubfigure}{\alph{subfigure}.2}
      \subfloat{\label{fig:HM_timeEvolution_J_2}}
      \addtocounter{subfigure}{-1}
      \renewcommand*{\thesubfigure}{\alph{subfigure}.3}
      \subfloat{\label{fig:HM_timeEvolution_J_3}}
      \renewcommand*{\thesubfigure}{\alph{subfigure}.1}
      \subfloat{\label{fig:HM_timeEvolution_E_1}}
      \addtocounter{subfigure}{-1}
      \renewcommand*{\thesubfigure}{\alph{subfigure}.2}
      \subfloat{\label{fig:HM_timeEvolution_E_2}}
      \addtocounter{subfigure}{-1}
      \renewcommand*{\thesubfigure}{\alph{subfigure}.3}
      \subfloat{\label{fig:HM_timeEvolution_E_3}}
      
      \includegraphics{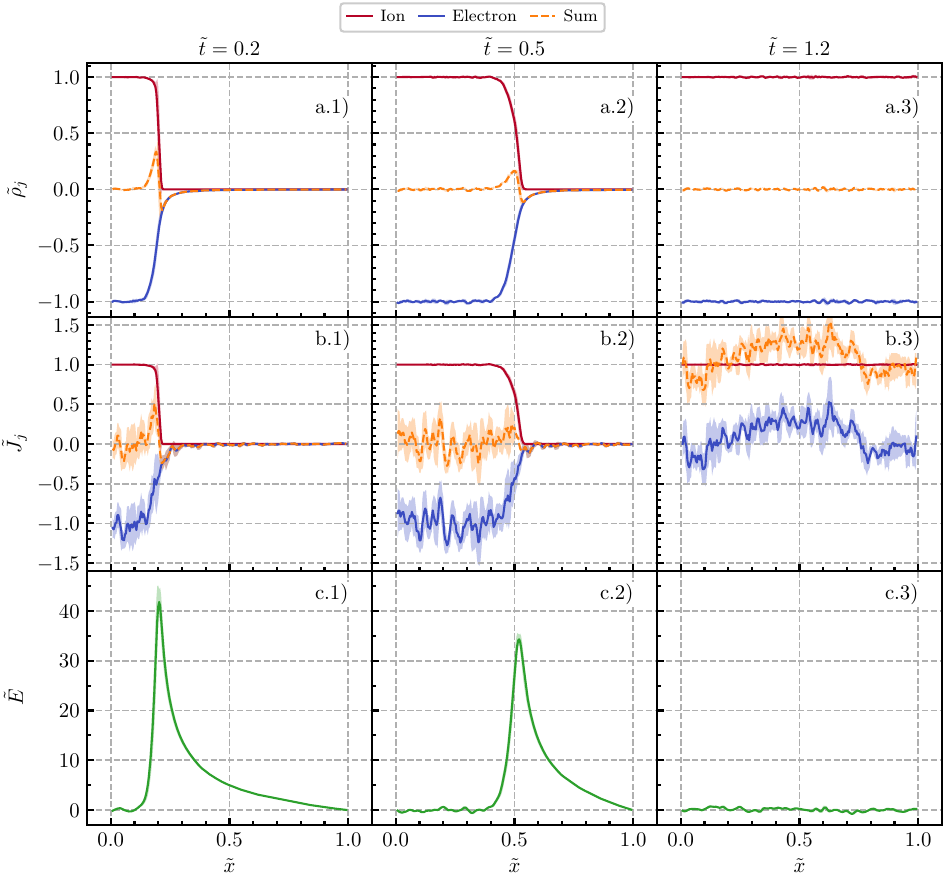}
      \caption{\label{fig:HM_timeEvolution}
               Time evolution of charge (top row) and current (middle row) densities of ions (red), electrons (blue), and the sum of both (orange) at three time steps $\tilde{t} = \numlist{0.2;0.5;1.2}$ (left, center, and right columns) for the case of injecting electrons with a half-Maxwellian distribution function and total reflection at the outflow boundary.
               The bottom row shows the electric field at each time.
               Data is smoothed with a moving average with a window length of $5$.
               Mean data is shown as a solid (dashed) line, and the standard deviation is shown as a shadowed region.
              }
    \end{figure*}
    
    Figure~\ref{fig:HM_timeEvolution} shows the evolution of charge density (top row), current density (middle row) and the electric field (bottom row) at three instants of the simulation, $\tilde{t} = \numlist{0.2;0.5;1.2}$.
    We see that even when we are injecting electrons with no drift velocity, \ie, $u_{e,0} = 0$, the plasma evolves into a situation of equal ion and electron currents, as seen in figures~\ref{fig:HM_timeEvolution_J_1} and~\ref{fig:HM_timeEvolution_J_2}.
    This indicates that only a fraction of electrons is returning due to the potential drop at the ion front, so that $J_e (v_x > 0) > J_e (v_x < 0)$.
    This, however, would result in a break of density quasi-neutrality, as fewer particles are returning to the influx boundary.
    
    \begin{figure}
      \centering
      \includegraphics{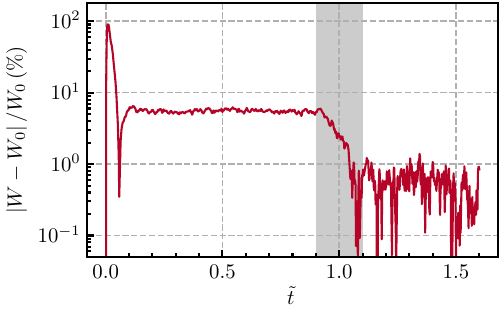}
      \caption{\label{fig:HM_inject}
               Relative weight of the injected electrons compared to the initial one.
               The value of the weight changes to maintain a zero electric field at the inflow.
               The gray shadowed region represents the time in which the ion front crosses the outflow boundary.
              }
    \end{figure}
    
    The simulation changes the weight of the injected particles with the correction scheme we set up at the injection, Sec.~\ref{sec:numericalSetup}, to adjust for the numerical noise and maintain quasi-neutrality.
    Figure~\ref{fig:HM_inject} shows the weight of injected electrons relative to the initial one.
    After some initial transitory region caused by the bulk injection of ions, $\tilde{t} < 0.1$, we see that the weight of injected electrons reaches \qty{60}{\percent} higher than the initial one, $\tilde{t} < 1$.
    This correction is necessary to ensure $n_e \simeq n_i$ as fewer electrons return to the injection plane in the free flowing plasma, meaning that the injected electrons have a higher weight to compensate.
    Once the ion front passes the outflow boundary, shadowed region in Fig.~\ref{fig:HM_inject}, all electrons are reflected back to ensure $J_e = 0$ as in the injection conditions.
    This results in a weight correction below \qty{1}{\percent}, enough to compensate for numerical noise.
    With these conditions for the electrons, half-Maxwellian injection and full reflection, the density quasi-neutrality is maintained in the quasi-steady state.
    However, we can appreciate how there is no quasi-neutrality of current, Fig.~\ref{fig:HM_timeEvolution_J_3}, as we have neglected the drift velocity of the electrons.
    
    This situation would be acceptable if a quasi-steady situation of the plasma were analyzed in which the accurate representation of the electron velocity is not required.
    However, this could affect situations in which an accurate electron energy distribution function is required, as collision processes, or when the self-consistent magnetic field needs to be considered.
    A correction to the outflow boundary condition could be implemented to maintain $\sum \vec{J}_j \simeq 0$,~\cite{li2019electron,andrews2022fully} but the conditions of the free-flowing plasma presented in Fig.~\ref{fig:HM_timeEvolution_J_2} point us to the need of injecting electrons with a drift velocity.
    
      
  \section{Injection of drift-Maxwellian electrons and outflow velocity control}\label{sec:driftMaxwellian}
    The injection of a drifted Maxwellian in which $u_{e,0} = u_{i,0}$, better represents the conditions reached by a flowing plasma, as seen in Fig.~\ref{fig:HM_timeEvolution_J_2}.
    This, however, presents an issue with the outflow boundary condition.
    In the previous case, Sec.~\ref{sec:haflMaxwellian}, a full reflection of the electrons was acceptable, as this would equal the density current imposed at the inflow and outflow boundaries in quasi-steady state.
    However, a total reflection in this case would lead to a zero current at the outflow, in disagreement with the inflow conditions.
    
    We have chosen to impose a boundary condition for the electrons that alters their reflection so their drift velocity matches that of the other charged species.
    This ensures that when total charge density is zero, \ie, $\sum \rho_j \simeq 0$, zero current density, $\sum \vec{J}_j \simeq 0$ is also guaranteed.
    
    \begin{figure}
      \centering
      \subfloat{\label{fig:boundaryScheme_a}}
      \subfloat{\label{fig:boundaryScheme_b}}
      \subfloat{\label{fig:boundaryScheme_c}}
      \includegraphics{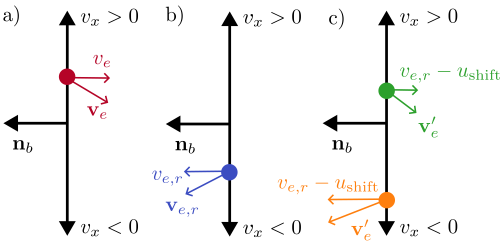}
      \caption{\label{fig:boundaryScheme}
               Reflection of the electrons to match the drift velocity of the remaining charged species in a 1D domain.
               The boundary has a normal vector $\vec{n}_b$ pointing inwards to the numerical domain.
               a) The electron (in red) collides with the boundary, and its velocity is projected normal to the surface, $v_e$.
               b) This velocity is then reflected as $v_{e,r}$ (in blue).
               c) Then, the electron velocity is shifted by $u_{\text{shift}}$.
               The resulting velocity can be positive (in green), in which case the particle escapes, or negative (in orange), in which case the electron is reflected into the numerical domain with velocity $\vec{v}_e^\prime$.
              }
    \end{figure}
    
    The reflection process is illustrated in Fig.~\ref{fig:boundaryScheme}, in which the normal vector of the boundary $\vec{n}_b$ points inwards to the numerical domain, which for the 1D case presented here is the negative $x$ direction.
    When an electron reaches the boundary, Fig.\ref{fig:boundaryScheme_a}, the velocity of the electron crossing the boundary projected on the boundary is $v_e$, which is positive as the particle is leaving the domain.
    The particle is reflected by the boundary (specular reflection), resulting in $v_{e,r}$, Fig.~\ref{fig:boundaryScheme_b}.
    The other components of the velocities are unaltered.
    Then, the velocity of an incident electron is shifted by $u_{\text{shift}}$ as
    \begin{equation}
       v_{e,r}^{\prime} = v_{e,r} - u_{\text{shift}}
    \end{equation}
    Two possibilities exist, as shown in Fig.~\ref{fig:boundaryScheme_c}.
    If the particle remains with a velocity pointing outwards from the domain, \ie, $v_{e,r}^{\prime} > 0$ it is removed from the simulation as it is considered to have enough energy to escape.
    Otherwise, the particle remains in the simulation with the new velocity achieved after the shift, $\vec{v}_e^\prime$.
    
    The shift velocity $u_{\text{shift}}$ is updated at each iteration so that
    \begin{equation}
        u_{\text{shift}}^{k+1} = u_{\text{shift}}^{k} + K \left( \vec{u}_e - \sum_{j \neq e} \vec{u}_j \right) \cdot \vec{n}_b
    \end{equation}
     where $k$ is the current iteration, $j$ refers to all the other charged species not being reflected, $\vec{u}_e$ is the drift velocity of electrons at the outflow boundary condition, and $\vec{n}_b$ is the normal vector to the edge pointing inwards to the numerical domain.
    The factor $K$ is a numerical smoothing value, typically set to \num{1e-1}, to avoid fast changes of $u_{\text{shift}}$ due to statistical noise.
    
    \begin{figure}
      \centering
      \subfloat{\label{fig:DM_timeEvolution_rho}}
      \subfloat{\label{fig:DM_timeEvolution_J}}
      \subfloat{\label{fig:DM_timeEvolution_E}}
      \includegraphics{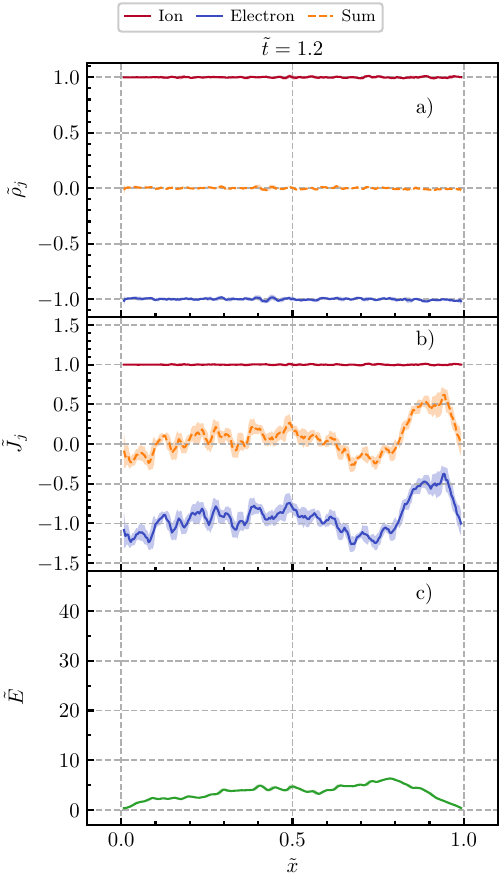}
      \caption{\label{fig:DM_timeEvolution}
               Plasma a) charge density, b) current density, and c) electric field at $\tilde{t} = \num{1.2}$ when injecting a drift Maxwellian distribution function so that the electron mean velocity equals the ion one and adjusting the outflow reflection of electrons to maintain quasi-neutrality in the current density.
              }
    \end{figure}
    
    When we inject an electron population as a drifted Maxwellian and adjust the reflection of particles at the outflow so that the total density current is zero, we reach a quasi-steady state in which quasi-neutrality of charge and current density is maintained, as seen in Fig.~\ref{fig:DM_timeEvolution}.
    The quasi-neutrality of charge density and electric field remains as in the case in Fig.~\ref{fig:HM_timeEvolution}, up to noise levels related to the simulation.
    But now, the current density quasi-neutrality, Fig~\ref{fig:DM_timeEvolution_J}, is also maintained, as the total value in the quasi-steady state oscillates around $0$.
    
    \begin{figure}
      \centering
      \subfloat{\label{fig:DM_outflow_u}}
      \subfloat{\label{fig:DM_outflow_ushift}}
      \includegraphics{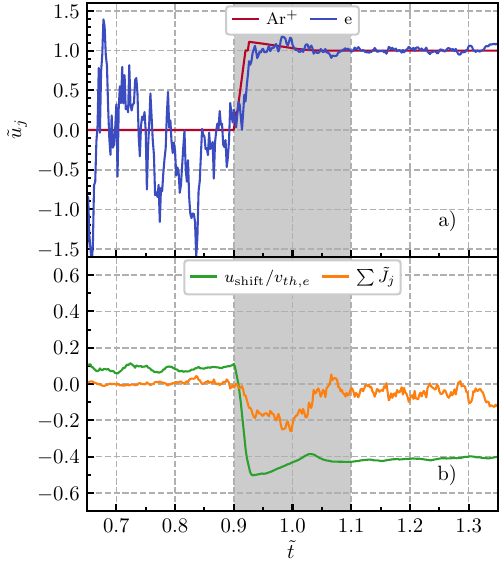}
      \caption{\label{fig:DM_outflow}
               a) Drift velocity of ions (red) and electrons (blue) at the outflow boundary and b) shift velocity $u_{\text{shift}}$ divided by the thermal velocity at \qty{2}{\eV} (green) and the total density current $\sum \tilde{J}_j$ (orange) around $\tilde{t} = 1$.
               The gray shadowed region represents the time in which the ion front crosses the outflow boundary.
              }
    \end{figure}
    
    With this boundary condition, we are only modifying the reflection of electrons based on the conditions at the boundary, not requiring any input from the rest of the domain as in Refs.~\onlinecite{andrews2022fully,jambunathan2020self,difede2023magnetic,yang2024numerical,guaita2025electron}.
    High velocity electrons, $v_e > u_{\text{shift}}$, escape the numerical domain, while those with lower velocity are reflected, avoiding the pump instability.
    
    Additionally, during the transient of the ion front crossing the boundary, around $\tilde{t} = 1$, electron drift velocity remains quite close to the ion one, Fig.~\ref{fig:DM_outflow_u}, but the lack of quasi-neutrality ($n_i \neq n_e$ in the ion front) results in a density current imbalance, as shown by the orange line in Fig.~\ref{fig:DM_outflow_ushift}.
    Thus, this boundary reflection model represents a transient situation with a charge and current imbalance when the ion front crosses the boundary, as well as the quasi-neutral state reached in the quasi-steady state.
        
    However, this model still presents an issue.
    Before the ion front reaches the boundary, a full reflection is applied as $u_i = 0$.
    Although this is takes place before the ion front reaches the boundary, \ie, when the electron population is small, it still points out an incomplete representation of the electrons downstream the expanding plasma.
  
  \section{Using Ampere-Maxwell equation to impose the electric field at the outflow}\label{sec:Ampere}
    The implementation of the previous reflection model for the electrons allows for a quasi-steady situation in which $\sum \rho_j \simeq 0$ and $ \sum \vec{J}_j \simeq 0$, while preserving charge and current non-equilibrium while the ion front crosses the boundary.
    However, a zero electric field boundary condition was still imposed.
    We have seen how a charge non-equilibrium exists in the ion front as the plasma evolves freely in the domain, Fig.~\ref{fig:HM_timeEvolution_rho_2}, creating an intense electric field, Fig.~\ref{fig:HM_timeEvolution_E_2}.
    Thus, we need to implement a boundary condition for the Poisson's equation \eqref{eq:Poisson} that captures the evolution of the electric field as the ion front crosses the outflow boundary condition.
    
    To do so, we recall Ampere-Maxwell's law
    \begin{equation}
      \nabla \times \vec{B} - \mu_0 \sum \vec{J}_j = \mu_0 \varepsilon_0 \fpartial{\vec{E}}{t},\label{eq:AmpereMaxwellLaw}
    \end{equation}
     where $\vec{B}$ is the magnetic flux density vector and $\mu_0$ is the vacuum permeability.
    In the case of null $\vec{B}$ or, more accurately, when its curl is negligible, which covers many cases of plasma expansion, Eq.~\eqref{eq:AmpereMaxwellLaw} reduces to
    \begin{equation}
      \fpartial{\vec{E}}{t} = - \frac{1}{\varepsilon_0} \sum \vec{J}_j, \label{eq:AmpereMaxwellLaw_noB}
    \end{equation}
     which describes the evolution of the electric field in time based on the density current.
    Numerically, we can express this equation at the outflow boundary as
    \begin{equation}
        E_{b}^{k+1} = E_{b}^{k} - \frac{\tau}{\varepsilon_0} \sum \vec{J}_j \cdot \vec{n}_b, \label{eq:AmpereMaxwellLaw_num}
    \end{equation}
     where $E_{b}$ is the electric field normal to the boundary, $\tau$ is the time step, $k$ is the current iteration, $\vec{n}_b$ is the normal vector to the boundary, and the currents are evaluated at the boundary position.
    The electric field obtained from this equation can then be implemented as a non-zero Neumann boundary condition.
     
    Thus, we have a way to describe the temporal variation of the electric field based on the current density at a specific point.
    Of course, this means that we need a boundary condition for the electrons that results in a zero total current in a quasi-steady state while a current density non-equilibrium appears while the ion front crosses the boundary.
    Thankfully, the reflection model we have implemented for the electrons in the previous section fulfills these conditions.
    
    \begin{figure}
      \centering
      \includegraphics{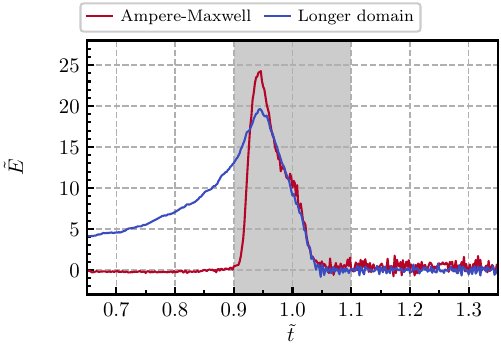}
      \caption{\label{fig:AM_boundariesEM}
               Electric field at the boundary when using the Ampere-Maxwell approach (red) and the electric field at the same position for a simulation with twice the domain length.
               The gray shadowed region represents the time in which the ion front crosses the outflow boundary.
              }
    \end{figure}
    
    We present in Fig.~\ref{fig:AM_boundariesEM} the electric field at $\tilde{x} = 1$ implementing this boundary condition compared to a case with twice the domain length, unperturbed by the boundaries.     
    We first notice a discrepancy in the electric field before the ion front crosses the boundary condition, $\tilde{t} = 1$, represented as a vertical dotted line.
    As the current at the boundary is close to zero, see Sec.~\ref{sec:driftMaxwellian}, the electric field remains close to its initial value, which for our initial boundary conditions is zero.
    This differs from the non-zero field found on the unperturbed case with the longer domain (blue line).
    The discrepancy clearly comes from the particle boundary imposed on the electrons, which affects the evolution of the electric field.
    
    However, as the ion front gets closer to $\tilde{x} = \num{1}$, we see the evolution of the unperturbed electric field following Eq.~\eqref{eq:AmpereMaxwellLaw_num}, red line.
    We especially notice an exact coincidence between the two cases starting at $\tilde{t} \simeq 0.95$.
    In the quasi-steady situation $\tilde{t} > 1.05$, both cases show a zero electric field, only perturbed by numerical noise.
    
    Thus, when we are dealing with a transient situation of current non-equilibrium, applying the Ampere-Maxwell's law to the outflow boundary condition represents the electric field of an unperturbed plasma crossing the boundary as well as the zero field at the quasi-steady state.
    Nevertheless, this does not represent properly the dynamic of the electron population downstream of the ion front.
    A better representation of the electron dynamics close to the boundary, coupled with the Ampere-Maxwell's law, could definitely improve the modeling of mesothermal plasma expansion.
      
  \section{Conclusions}\label{sec:conclusions}
    We have studied the effect of different injection conditions for the electrons in the fully kinetic simulation of mesothermal plasmas.
    When injecting electrons with a half-Maxwellian distribution, \ie, $u_{e,0} = 0$, the free flowing plasma reaches a situation in which $u_e \simeq u_i$ when unaffected by the boundary conditions.
    This results in a charge unbalance at the injection, as fewer electrons are reflected, which is solved numerically by adjusting the electron macro-particle to maintain a zero electric field.
    This correction was implemented to compensate for possible numerical noise and should only produce small changes in the weight compared to the initial one.
    However, while the plasma is flowing through the domain, a strong correction is required, as shown in Fig.~\ref{fig:HM_inject}.
    If we impose a full reflection of electrons at the outflow to maintain $u_e = 0$, the weight goes back to the initial one once the ion front has crossed the outflow boundary.
    However, this leads to a quasi-steady state in which current quasi-neutrality, $\sum \vec{J}_j \simeq 0$, is violated.
    This approach could be acceptable for studying quasi-steady states in which the electron current is not relevant, such as collisionless plasmas or cases that neglect the self-consistent magnetic field.
    Of course, different outflow boundary conditions could be applied to obtain a non-zero electron current while maintaining the injection of half-Maxwellian electrons, but the need to use a strong correction at injection to ensure quasi-neutrality could lead to issues in the electron energy distribution function.
    
    Thus, we have proposed to inject the electrons with a drift-Maxwellian distribution function so $u_{e,0} = u_{i,0}$.
    Of course, this means we cannot impose full reflection of the electrons at the outflow boundary.
    We have implemented a boundary condition that adjusts the reflection of electrons so that the drift velocity of those reaching the outflow boundary matches that of the ions.
    This approach describes transitory states, as when the ion front crosses the boundary, but also ensures current quasi-neutrality in a quasi-steady state.
    Nevertheless, we have also noticed that when no ions reach the boundary, this condition behaves as a full reflection and does not properly represent the dynamic of electrons downstream of the ion front.
    
    These previous cases impose a zero-field Neumann condition at the outflow boundary, \ie, $\ipartial{\phi}{x} = 0$ in our numerical domain.
    However, the ion front presents a charge density disequilibrium and thus a non-zero electric field.
    We propose to use Ampere-Maxwell's law to calculate a time evolving electric field based on the density current crossing the boundary.
    We have shown that we can reproduce the electric field of the unperturbed solution, Fig.~\ref{fig:AM_boundariesEM}, although some issues remain when only the electrons downstream of the ion front exist at the boundary.
    This is also related to the way we have decided to reflect electrons.
    Improving the particle boundary for electrons when no ions are present would definitely contribute to a description of the electric field closer to the unperturbed plasma.
    Additionally, the fact that we require an electrodynamic field at the boundary indicates that we should not assume an electrostatic field for a flowing mesothermal plasma, and a more complete solution of Maxwell's equations is required.
    
  \begin{acknowledgments}
    I would like to thank Dr. J. M. Donoso and (soon to be Dr.) I. Gomez for their helpful comments while writing this manuscript.
  
  \end{acknowledgments}
  
  \bibliography{bibliography}

\end{document}